\documentclass[MSNbibl,nameyear,dvips]{arxstspdf}
\usepackage{flushend}
\usepackage{stfloats}
\usepackage{url,breakurl}

\volume{29}
\issue{1}
\pubyear{2014}
\firstpage{98}
\lastpage{100}
\doi{10.1214/13-STS446} 

\begin{document}
\begin{frontmatter}

\title{Wonderful Examples, but Let's not Close Our Eyes}
\runtitle{Wonderful Examples}

\begin{aug}
\author[a]{\fnms{David J.} \snm{Hand}\corref{}\ead[label=e1]{d.j.hand@imperial.ac.uk}}
\runauthor{D. J. Hand}

\affiliation{Imperial College of Science, Technology, and Medicine}

\address[a]{David Hand is Emeritus Professor of Mathematics, Department of Mathematics,
Imperial College of Science, Technology, and Medicine, Huxley Building, 180
Queen's Gate, London SW7 2BZ, United Kingdom \printead{e1}.}

\end{aug}

\begin{abstract}
The papers in this collection are superb illustrations of the power of
modern Bayesian methods. They give examples of problems which are well
suited to being tackled using such methods, but one must not lose sight of
the merits of having multiple different strategies and tools in one's
inferential armoury.
\end{abstract}

\begin{keyword}\hspace*{-1.5pt}
\kwd{Frequentist}
\kwd{likelihood inference}
\kwd{Neyman--Pearson hypothesis testing}
\kwd{schools of inference}
\end{keyword}

\end{frontmatter}

Space prohibits me from making specific comments on each of these
informative and thought-provoking papers---they each merit an extended
discussion in their own right. Instead, I will make some general comments
about the collection.

The papers provide marvelous examples of the power of modern statistics
and, in particular, of the power of modern Bayesian methods. The adjective
``modern'' here is intended mainly to indicate that it is the power of the
computer which has made practical solutions such as those illustrated in
these papers. But I~have to ask, is the emphasis on ``Bayesian'' necessary?
That is, do we need further demonstrations aimed at promoting the merits of
Bayesian methods? Surely the case is proven: Bayesian methods are very well
suited to tackling many problems, leading to solutions which would be hard
to arrive at by alternative methods.

The examples in this special issue were selected first by the authors, who
decided what to write about, and, then, second by the editors, in deciding
the extent to which the articles conformed to their desiderata of being
Bayesian success stories: that they ``present actual data processing
stories where a non-Bayesian solution would have failed or produced
suboptimal results.'' In a way I think this is unfortunate. I am certainly
convinced of the power of Bayesian inference for tackling many problems,
but the generality and power of the method is not really demonstrated by a
collection specifically selected on the grounds that this approach works
and others fail. To take just one example, choosing problems which would be
difficult to attack using the Neyman--Pearson hypothesis testing strategy
would not be a convincing demonstration of a weakness of that approach if
those problems lay outside the class that approach was designed to
attack. One of the basic premises of science is that you must not select
the data points which support your theory, discarding those which do not.
In fact, on the contrary, one should \textit{test} one's theory by
challenging it with tough problems or new observations. (This contrasts
with political party rallies, where the candidates speak to a cheering
audience of those who already support them.) So the fact that the articles
in this collection provide wonderful stories illustrating the power of
modern Bayesian methods is rather tarnished by the one-sidedness of the
story. If I wasn't already convinced of the power of the Bayesian paradigm,
I might be tempted to wonder if there was too much protestation going on.

Or perhaps, if one is going to have a collection of papers demonstrating
the power of one particular inferential school, then, in the journalistic
spirit of balanced reporting, we should invite a series of similar
 articles which ``present actual data processing stories where a
nonfrequentist/nonlikelihood/non-[fill in your favorite school of
inference] solution would have failed or produced suboptimal results.'' Or
even examples of the power of each of the \textit{other} 46655 different
varieties of Bayesian approach (Good, \citeyear{Goo71}).

The editors emphasized that they were not looking for ``argumentative
rehashes of the Bayesian versus frequentist debate.'' I can only commend
them on that. On the other hand, times move on, ideas develop, and
understanding deepens, so while ``argumentative rehashes'' might not be
desirable, reexamination from a more sophisticated perspective might be.
The editors went on to say ``we the editors are convinced of the generic
appeal of `doing it Bayes' way,' while non-Bayesians are convinced of the
opposite.'' I think this is a slightly unfortunate phrasing. I would
(admittedly, perhaps naively) like to think that any modern statistician
would look at each problem on its merits, and decide what ``way'' was best
suited to tackle that problem. I am always a little uncomfortable when I
hear about ``the one true way'' of looking at things.

An interesting question, perhaps in part sociological, is why different
scientific communities tend to favor different schools of inference.
Astronomers favor Bayesian methods, particle physicists and psychologists
seem to favor frequentist methods. Is there something about these different
domains which makes them more amenable to attack by different approaches?

In general, when building statistical models, we must not forget that the
aim is to understand something about the real world. Or predict, choose an
action, make a decision, summarize evidence, and so on, but always about
the real world, not an abstract mathematical world: our models are not the
reality---a point well made by George Box in his oft-cited remark that
``all models are wrong, but some are useful'' (Box, \citeyear{autokey1}). So, likewise, if
different models suit different purposes, why should we expect one approach
to inference to be universally applicable? The internal mathematical
coherence of Bayesian methods is very attractive, but it must not be
allowed to take priority over the ultimate aim, which is to say something
about the reality we are studying. As Albert Einstein put it: ``as far as
the propositions of mathematics refer to reality, they are not certain; and
as far as they are certain, they do not refer to reality.'' (Einstein,
\citeyear{Ein21}). As an aside, there is also the question of what exactly is meant by
``Bayesian.''\vadjust{\goodbreak}
Cox and Donnelly [(\citeyear{CoxDon11}), page 144] remark that ``the word
\textit{Bayesian}, however, became ever more widely used, sometimes
representing a regression to the older usage of `flat' prior distributions
supposedly representing initial ignorance, sometimes meaning models in
which the parameters of interest are regarded as random variables and
occasionally meaning little more than that the laws of probability are
somewhere invoked.''

Turning to the papers themselves, the Bayesian approach to statistics, with
its interpretation of parameters as random variables, has the merit of
formulating everything in a consistent manner. Instead of trying to fit
together objects of various different kinds, one merely has a single common
type of brick to use, which certainly makes life easier. In particular,
this means that very elaborate models can be handled with relative ease. As
is elegantly demonstrated in the papers, although the model formulation
requires deep and careful thought, at some level the Bayesian procedure is
attractively straightforward.

On the basis of these papers, one can certainly see the sorts of problems
which lend themselves to attack by Bayesian methods and which are difficult
to approach in other ways. Common characteristics seem to be complex
models, fragmentary and indirect evidence, the task being evidence
synthesis or explicitly to develop a probability distribution, and so on.
Each of these are tough problems to cope with, and one should be reassured
that statisticians now have the tools to tackle them.

But reassurance should not drift into complacency. When presented with
fragmentary evidence, for example, one should proceed with caution. In such
circumstances, the opportunity for undetected selection bias is
considerable. Assumptions about the missing data mechanism may be
untestable, perhaps even unnoticed. Data can be missing only in the context
of a larger model, and one might not have any idea about what model might
be suitable. Having an inferential strategy which can cope with such
problems should not tempt one to ignore the fact that they are there, along
with the consequent qualifications and reservations about the conclusions
drawn.

Likewise, the power of Bayesian methods to handle complex models is very
exciting. Many problems statisticians are asked to tackle are complex, and
a complex model is necessary. So the fact that we statisticians now have a
paradigm which will allow us to tackle increasingly complex models is
certainly to\vadjust{\goodbreak} be applauded. But I do have this nagging feeling that
sometimes a more approximate solution might be more suitable. On the one
hand, very elaborate models have many ways to be misspecified, and, on the
other, statisticians rarely work on practical problems in isolation, but
typically in conjunction with domain experts in the area being explored.
The statistician brings statistical expertise, but at the end of it all the
answer must be comprehensible to the other scientists: one aspect of a
model being ``useful,'' to use Box's word, is that it  should be
comprehensible. And there are other related aspects. Timeliness, a
corollary of simplicity, is one. I~am reminded of a comment made by David
Lawrence of Citicorp: ``In one business, we waited more than 20 months for
a professor of statistics to come up with the `Cadillac' of scoring
systems, while all the business needed was a `Chevrolet' that would
work.'' (Lawrence, \citeyear{Law84}, page 55).

You will see that I am trying to argue the case of balance. Despite that,
and however you look at it, the editors are to be congratulated on
collating a superb collection of papers illustrating the power of modern
statistics to handle complex problems. Moreover, within the remit of what
they set out to do---to demonstrate the power of modern Bayesian
methods---they
certainly succeeded. I shall definitely draw the attention of my
students to this excellent collection of articles.




\end{document}